# A note on an invariant for one-dimensional heat conduction

Hamou Sadat, Christian Prax and Vital Le Dez

Institut PPRIME, UPR CNRS 3346, Université de Poitiers, 40 Avenue du Recteur Pineau, 86022 Poitiers, France

In a recent article [1], an invariant for the one dimensional heat conduction has been introduced. When temperature is lower than the Debye temperature, this invariant reduces to:

$$T^4(z) + T^4(L-z) = T^4(0) + T^4(L), \quad (1)$$

where z is the abscissa along the one-dimensional rod of length L and T is temperature. This result has been established by solving one-dimensional Boltzmann transport equation and by assuming that equilibrium phonon intensity is proportional to the fourth power of temperature (see equation 2 therein). In the same article, experimental results have also been carried out and apparently confirmed the correctness of the statement. It is to notice that the experiment has been carried out with a rod of total length L~98mm with one end maintained at low temperature by using a cold bath (T~200K) and with an imposed flux of order 100 kW/m² at the other end. At this macroscopic scale, heat conduction can therefore be accurately modeled by the Fourier law. The main goal of this note is therefore to show that in steady heat conduction, the proposed invariant given by equation (1) is nothing than an approximate solution to the Fourier solution (in the conditions of the experiment).

Let us first consider the case of steady heat conduction in the copper rod examined in [1]. In the temperature range of the experiment (190K-270K), the thermal conductivity is nearly constant (k~395 W/m K) and the steady heat conduction solution is the well-known linear temperature variation. One can therefore write that $T_2 = T_3 + \alpha$ and $T_4 = T_3 - \alpha$ where: $\alpha = \frac{q\,d}{\lambda}$ is the temperature difference in steady state. One can also write the well-known relation: $T_2 + T_4 = 2T_3$. We can now calculate $T_2^4$ and $T_4^4$ and write:

$$T_2^4 + T_4^4 = 2T_3^4 + 12 T_3^2 \alpha^2 + 2\alpha^4, \quad (2)$$

which gives (for values of $\alpha$ much smaller than temperature):

$$T_2^4 + T_4^4 \sim 2T_3^4 \quad (4)$$

For example, with the conditions of the experiment (a flux of order of 100 kW/m², a distance d=22.9 mm and a temperature around 230 K), the last two terms are of equation (2) are: $12T_3^2\alpha^2 = 2.07\ 10^7\ K^4$ and $\alpha^4 = 1070.5\ K^4$. These two values are negligible when compared to: $2T_3^4 = 5.59\ 10^9\ K^4$. We now consider a material whose conductivity varies with the inverse of temperature: $k = \frac{A}{T}$, like silicon for example. The solution of the steady heat equation is: $T = T_0 e^{\frac{x}{L}ln\left(\frac{T_L}{T_0}\right)} = T_0 \left(\frac{T_L}{T_0}\right)^{\frac{x}{L}}$.

We hence have:

$$T^4(x) + T^4(L-x) = T_0^4\left[\left(\frac{T_L}{T_0}\right)^{\frac{4x}{L}} + \left(\frac{T_L}{T_0}\right)^{\frac{4(L-x)}{L}}\right] \quad (5)$$

It can be shown that this can be written:

$$T^4(x) + T^4(L-x) = 2T_0^4\left(\frac{T_L}{T_0}\right)^2 cosh\left[\frac{2L-4x}{L}ln\left(\frac{T_L}{T_0}\right)\right] \quad (6)$$

When x=L/4, this reduces to:

$$T_2^4 + T_4^4 = 2T_0^4\left(\frac{T_L}{T_0}\right)^2 cosh\left[ln\left(\frac{T_L}{T_0}\right)\right] = 2T_3^4 cosh\left[ln\left(\frac{T_L}{T_0}\right)\right] \quad (7)$$

If we take a temperature variation in the silicon rod of 30K and a lower temperature at 200K, we get: $cosh\left[ln\left(\frac{T_L}{T_0}\right)\right]$=1.0098 and once again, one is lead to approximate equation (4). These two examples show clearly that the proposed invariant is nothing than an approximate solution to the steady heat equation.

We now turn to the experimental results presented in [1]. It is first worth to note that radiative loss from the surface of the rods was estimated by the authors to be 'smaller than 1% of the conductive heat flux along them'. The authors also claim that steady state-state has been reached. Therefore, the problem is that of one dimensional steady heat conduction. In this condition, results gathered on Figures 4 are surprising. This can be verified by calculating the heat flux when using for example the experimental point on Figure 4a corresponding to copper with (T3-T4)=7K and (T2-T3)=5K. Two different values of the flux, namely q=122 kW/m² and q=87.3 kW/m², are obtained and energy conservation principle is not fullfilled. In our opinion, this difference on the calculated heat flux shows that the rod was probably not yet in steady state equilibrium during the measurements. It remains to explain why the experimental temperatures fit quite well on a 45° line when plotted in the T⁴ form in Figure 5. Toward this end, let us suppose that we have : $T_2 = T_3 + \frac{q_2 d}{k}$ and $T_4 = T_3 - \frac{q_4 d}{k}$ and calculate the temperature fourth powers as follows:

$$T_2^4 = T_3^4 (1 + \frac{q_2 d}{k T_3})^4 \quad \text{and} \quad T_4^4 = T_3^4 (1 - \frac{q_4 d}{k T_3})^4 \quad (8)$$

We have seen that in the experiments carried out in [1], $\frac{q_2 d}{k T_3}$ and $\frac{q_4 d}{k T_3}$ are small quantities. Hence, we can write:

$$T_2^4 \sim T_3^4 (1 + 4\frac{q_2 d}{k T_3}) \quad \text{and} \quad T_4^4 = T_3^4 (1 - 4\frac{q_4 d}{k T_3}) \quad (9)$$

This leads to:

$$T_2^4 + T_4^4 \sim 2 T_3^4 + 4 \frac{d}{k} T_3^3 (q_2 - q_4) \quad (10)$$

The last term of the right hand of previous equation is negligible compared to the first. With the calculated fluxes q=122kW/m² and q=87.3kW/m² and a temperature $T_3$=230K, one obtains for example: $4 \frac{d}{k} T_3^3 (q_2 - q_4) = 9.8 \; 10^7 K^4$ which is to be compared to $2 T_3^4 = 5.6 \; 10^9 \; K^4$. The ratio of the two terms is equal to 1.7% and therefore the approximate solution given by equation (4) holds (at 1.7%).

We conclude that equation (1) does not hold in macroscopic scale heat conduction problems although it remains a good approximation in some particular situations such as in the experiment of [1]. The proposed invariant is nothing than an approximation to the Fourier solution at macroscopic scales.